\documentclass[12pt]{article}
\usepackage{amssymb}
\usepackage{amsmath}
\usepackage{amsthm}
\usepackage{latexsym}
\usepackage{mathdots}
\usepackage{graphicx}

\newtheorem{prop}{Proposition}
\newtheorem{construction}{Construction}

\newtheorem{rem}{Remark}
\newtheorem{definition}{Definition}
\newtheorem{example}{Example}

\newcommand{\cC}{{\mathcal C}}
\newcommand{\cD}{{\mathcal D}}
\newcommand{\cG}{{\mathcal G}}
\newcommand{\cH}{{\mathcal H}}

\newcommand{\cM}{{\mathcal M}}

\newcommand{\cT}{{\mathcal T}}

\newcommand{\cW}{{\mathcal W}}
\newcommand{\cZ}{{\mathcal Z}}

\newcommand{\T}{\mbox{$\!^{\sf T}$}}

\newcommand{\dd}{\mbox{\rm d}}

\newcommand{\GL}{\mbox{\rm GL}}
\newcommand{\wt}{\mbox{\rm wt}}

\newcommand{\spann}{{\rm span}}

\newcommand{\im}{{\rm im}\,}

\newcommand{\GRS}{{\rm GRS}}

\newcommand{\Smallfourmat}[4]{\mbox{$\left(\begin{smallmatrix}{#1}&{#2}\\{#3}&{#4}\end{smallmatrix}\right)$}}

\newenvironment{liste}{\begin{list}{$\bullet$\hfill}{\labelwidth.4cm
   \topsep0ex \leftmargin.5cm \labelsep.1cm \rightmargin0cm \parsep0ex \itemsep0ex
   \partopsep0ex}}{\end{list}}
\newcounter{alp}
\newcounter{ara}
\newcounter{rom}

\newenvironment{alphalist}{\begin{list}{(\alph{alp})\hfill}{\usecounter{alp}
     \topsep0ex \labelwidth.7cm \leftmargin.7cm \labelsep0cm
     \rightmargin0cm \parsep0ex \itemsep.5ex
     \partopsep0ex}}{\end{list}}
\newenvironment{arabiclist}{\begin{list}{(\arabic{ara})\hfill}{\usecounter{ara}
     \topsep0ex \labelwidth.7cm \leftmargin.7cm \labelsep0cm
     \rightmargin0cm \parsep0ex \itemsep.5ex
     \partopsep0ex}}{\end{list}}

\usepackage{amsmath, amsfonts}
\usepackage{mathptmx}       
\usepackage{helvet}         
\usepackage{courier}        
\usepackage{type1cm}        
\usepackage{makeidx}         
\usepackage{graphicx}        
\usepackage{multicol}        
\usepackage[bottom]{footmisc}

\makeindex             

\begin{document}

\title{Variations of the McEliece Cryptosystem}
\author{Jessalyn Bolkema, Heide Gluesing-Luerssen, Christine A. Kelley, \\
  Kristin Lauter, Beth Malmskog, Joachim Rosenthal \footnote{\mbox{} Jessalyn Bolkema at University of Nebraska--Lincoln, {jessalyn.bolkema@huskers.unl.edu}; Heide Gluesing-Luerssen at University of Kentucky, {heide.gl@uky.edu};
Christine Kelley at University of Nebraska--Lincoln, {ckelley2@math.unl.edu};
Kristin Lauter at Microsoft Research, {klauter@microsoft.com};
Beth Malmskog at Vilanova University, {beth.malmskog@gmail.com}; Joachim Rosenthal at University of Z{\"u}rich, {rosenthal@math.uzh.ch}}
}

%
%
\maketitle

\abstract{Two variations of the McEliece cryptosystem are
presented.  The first is based on a relaxation of the column
permutation in the classical McEliece scrambling process.  This is done in such a way
that the Hamming weight of the error, added in the encryption process,
can be controlled so that efficient decryption remains possible.  The
second variation is based on the use of spatially coupled
moderate-density parity-check codes as secret codes.  These codes are
known for their excellent error-correction performance and allow for a
relatively low key size in the cryptosystem.  For both variants the
security with respect to known attacks is discussed.}

\section{Introduction}
Many widely-used public key cryptosystems, such as RSA and elliptic
curve cryptography, can be broken by a large-scale quantum computer.
It was pointed out in a report by National Institute of Standards and
Technology~\cite{nist16} that the building of powerful quantum
computers might well be feasible in a matter of 20 years and for this
reason the report encourages academics to come up with systems which
might still be safe in an environment where quantum computers exist.

One promising family of public key systems which are potentially secure in a
post-quantum computer environment are so called code-based
cryptosystems, like the McEliece~\cite{McE78} and the
Niederreiter~\cite{Nie86} systems.  These systems are based on the
hardness of decoding a random linear code, a problem that has so far
proved resistant to quantum computer-aided attacks.  In both these
cryptosystems the secret key is a code for which an efficient decoding
algorithm is known.  The public key is a disguised version of the
secret code that appears to be a random code, thus hiding the structure
that gives rise to efficient decoding.

The security of such a system is thus based on two assumptions.
First, it is intractable to decode the seemingly random public code with 
general decoding algorithms.  Since the best known algorithms for
decoding random linear codes are exponential in the length of the
code~\cite{MMT11,BJMM12}, this can in principle be achieved by
increasing the size of the code.  However, this comes at the cost of a
prohibitively large key size.  The second assumption is that it is
impossible for an attacker to reconstruct or uncover the underlying
structure of the secret code from the public code.  This prevents the
attacker from simply using the efficient decoding algorithm herself.

The second assumption has held for McEliece's original proposal, but
has been proven wrong for many later variants.  McEliece~\cite{McE78}
suggests a binary Goppa code as the efficiently decodable secret key.
The code is disguised through a random column permutation of its
generator matrix and left multiplication by a random invertible
matrix.  Encryption consists of encoding a plaintext message with the
public code and adding a random error whose Hamming weight stays
within the error-correcting bound of the secret code (which is the
same as the bound of the public code).  An attacker would then have to
correct this error in order to recover the plaintext.  For a precise
description of the McEliece cryptosystem based on arbitrary
efficiently decodable codes, see Section~\ref{S-McENie} of this paper.

To this day, the McEliece system remains unbroken -- if its originally
suggested size is adjusted in order to defend against current computer
power and incremental speed-ups in decoding algorithms~\cite{BLP08}.
However, its major disadvantage is the large public key size of the
system.  For this reason many attempts have been made to find
alternatives to McEliece's original proposal using different
underlying codes equipped with more compact representations,
including both codes with algebraic structure like Reed-Solomon codes
and modern codes like low-density parity-check codes with
quasi-cyclic structure.


In (almost) all these systems the masking of the secret code consists
of a random permutation and rescaling of the columns of a generator
matrix along with left multiplication by an invertible matrix.  While
the left multiplication is clearly just a change of basis of the code,
the column operations lead to a so-called monomially equivalent code.
Since permuting and rescaling of codeword coordinates leave the
Hamming weight invariant, these operations can easily be dealt with
when setting up the cryptosystem.  On the downside, this type of
masking often leaves too much algebraic structure, which an attacker
can exploit.  Indeed, in most cases the algebraic type of the code
remains unchanged: for example, a disguised generalized Reed-Solomon
code is a generalized Reed-Solomon code.  As a consequence, the
knowledge of the code class provides an attacker with further useful
information.  We give an overview of many proposed variants and
attacks in Remark~\ref{R-Insecure}.  The only codes that appear to
resist such an attack are alternant codes~\cite{BLP08}.
In Section~\ref{S-Wt2} we
elaborate on the idea of Baldi et al.~\cite{BBCRS16} by proposing a
McEliece cryptosystem where the monomial transformations are replaced
by the inverses of (invertible) matrices whose rows have Hamming
weight~$m=2$.  This generalization appears to have the advantage of
annihilating the algebraic structure of the secret code without
leaving a trace for an attacker to restore it using the distinguisher
coming from the Schur square of the public code.  Indeed, our
simulations show that the Schur square of the public code, a central
tool in most attacks, behaves like the Schur square of a random code.
This contrasts the system proposed by Baldi et al.~\cite{BBCRS16} for
which it has been shown~\cite{co15a} that the dimension of the Schur
square is not maximal when the average row weight $m$ is less than
$1+R$.

As an alternative to improved masking in algebraic systems,
in Section~\ref{S-SCMDPC} we
propose the use of spatially coupled Moderate-Density Parity-Check
(SC-MDPC) codes for the McEliece cryptosystem. Low-Density
Parity-Check (LDPC) codes were first considered as candidates in
\cite{mo00p}, but were soon observed to be vulnerable to decoding
attacks due to the high density of low-weight codewords in the dual code; a brute force search for such codewords allows attackers to generate a low-density parity-check matrix to
efficiently decode the code.  Furthermore, general LDPC-based McEliece
cryptosystems suffer from large key sizes, motivating quasi-cyclic
LDPC (QC-LDPC) codes to be considered~\cite{ba07p,BBC08}.  For
increased security, Moderate-Density Parity-Check (MDPC) codes were
considered in \cite{MTSB13}, and in particular, quasi-cyclic MDPC
(QC-MDPC) codes. Most recently a new attack on all classes of quasi-cyclic MDPC codes has appeared by Guo et al.~\cite{gu16p}, suggesting that the use of quasi-cyclic structure to reduce key size introduces a weakness in security. Since spatially coupled codes
are known for their efficient decoding, capacity approaching performance,
minimum distance properties, and
compact representation \cite{cdfkms14,mld15}, the use of these codes offers a promising
variant of the cryptosystem. In particular, spatially coupled codes offer a reduced key size without the security cost of rigid quasi-cyclic structure.

In order to check the vulnerability of our proposed systems with respect to
brute-force decoding, that is, information-set decoding, we compute
the work factor for Stern's algorithm for some sets of parameters and
secret generalized Reed-Solomon (GRS) codes in Section~\ref{S-Wt2}. Taking the reduced
Hamming weight of the errors into account, we show that the system can
be made safe with respect to information-set decoding.  Though the
keys are still large, our variation compares favorably with current
post-quantum alternatives. Similarly, in Section~\ref{S-SCMDPC}, we study the work factor
for classical decoding attacks and key recover attacks of the spatially coupled MDPC system.
In both cases, our variation compares favorably to comparable quasi-cyclic systems.

\section{Coding-Theoretic Preliminaries}\label{S-Basics}
Let us recall some basic notions of coding theory.  Throughout
let~$\mathbb{F}=\mathbb{F}_q$ be a finite field with~$q$ elements. We endow $\mathbb{F}^n$
with the Hamming weight $\wt(v_1,\ldots,v_n):=|\{i\mid
v_i\neq0\}|$. The associated distance is denoted by $\dd$, thus
$\dd(v,w):=\wt(v-w)$.  An $[n,k]_q$-code $\cC$ is a $k$-dimensional
subspace of the metric space $(\mathbb{F}^n,\dd)$.  The minimum distance
of~$\cC$ is defined as $\dd(\cC)=\min\{\wt(v)\mid
v\in\cC\backslash\{0\}\}$, and we call~$\cC$ an $[n,k,d]_q$-code
(resp.\ $[n,k,\geq d]_q$-code) if it is an $[n,k]_q$-code with
distance~$d$ (resp.\ at least~$d$).  Any $[n,k]_q$-code~$\cC$ can be
written in either of the forms
\[
\cC=\im(G):=\{uG\mid u\in\mathbb{F}^k\}\ \text{ and }\
\cC=\ker(H):=\{v\in\mathbb{F}^n\mid Hv\T=0\}
\]
for some matrix $G\in\mathbb{F}^{k\times n}$ (whose rows thus form a basis
of~$\cC$) and some $H\in\mathbb{F}^{(n-k)\times n}$.  We call any such
matrix~$G$ a \emph{generator matrix} and~$H$ a \emph{parity check
  matrix} of~$\cC$.  Define the dual of $\cC\subseteq\mathbb{F}^n$ as
$\cC^{\perp}:=\{w\in\mathbb{F}^n\mid v\cdot w=0\text{ for all }v\in\cC\}$,
where $w\cdot v$ is the standard dot product. Clearly, if
$\cC=\im(G)=\ker(H)$ as above, then $\cC^{\perp}=\im(H)=\ker(G)$.

Let~$\cC$ be an $[n,k,\geq d]_q$-code.  The triangle inequality of the
Hamming metric shows that for any $z\in\mathbb{F}^n$ satisfying
$\dd(z,\cC):=\min\{\dd(z,v)\mid v\in\cC\}\leq\frac{d-1}{2}$ there
exists a unique codeword $v\in\cC$ such that
$\dd(v,z)\leq\frac{d-1}{2}$.  Only codes where this unique codeword
can be found efficiently are suitable for error-correcting purposes.
There exist various classes of algebraic codes with efficient decoding
algorithms, such as generalized Reed-Solomon codes, BCH-codes,
alternant codes, Reed-Muller codes, AG-codes. Moreover, graph-based
codes are equipped with low-complexity iterative decoders.

Code-based cryptography relies on the fact that decoding up to
$(d-1)/2$ errors with respect to a random linear code is NP-hard;
see~\cite{BMT78}.  Furthermore, the best known decoding algorithms for
a random linear code, known as information-set decoding, have time
complexity that is exponential in the length and rate.

One such algorithm is due to Stern~\cite{St89}, with a precursor given by Lee/Brickell in~\cite{LeBr89}.
Stern's algorithm aims at finding error vectors with weights
$p,\,p,\,0$ when restricted to certain index sets $X,\,Y,\,Z$,
respectively.  Here~$X$ and~$Y$ are sets of size $k/2$, and~$Z$ has
size~$\ell$.  
The parameters~$p$ and~$\ell$ need to be chosen as to optimize the average run time.
A detailed description of the algorithm for codes over arbitrary finite fields can be found in~\cite[Sec.~3]{Pe10} 
by Peters.
In the same paper, Peters also presents a work factor estimate.
For $\mathbb{F}=\mathbb{F}_q$ with $q=2^m$ and the chosen parameters~$p$ and~$\ell$ this results in
\begin{equation}\label{e-WFStern}
  \tilde{W}^{\text{bin}}_{p,\ell}=m S_{p,\ell}{n\choose r}\bigg[{k/2\choose p}^2{n-k-\ell\choose r-2p}\bigg]^{-1}
  \ \text{ binary operations,}
\end{equation}
where
\begin{align*}
  &S_{p,\ell}:=(n-k)^2(n+k)+\bigg[k/2-p+1+2{k/2\choose p}(q-1)^p\bigg]\ell\\[.7ex]
  &\mbox{}\qquad\quad\
  +\frac{2pq(r-2p+1)(2q-3)(q-1)^{2p-2}}{q^\ell}{k/2\choose p}^2.
\end{align*}
The parameters~$p$ and~$\ell$ have to be determined to minimize $\tilde{W}^{\text{bin}}_{p,\ell}$.  
The effort can be further reduced by making wise use of computations in a previous round of the iterations.
We refer to~\cite{BLP08} by Bernstein et al.\ and~\cite{Pe10} by Peters for
further details and remark that this re-using of earlier computations
is an extension of an original improvement by Canteaut/Chabaud~\cite{CaCh98}.  
An additional improvement of Stern's algorithm has been proposed by Finiasz/Sendrier~\cite{FiSe09};
see also \cite[Sec.~6]{Pe10} for arbitrary finite fields.  
It is interesting to note that in \cite{CantoTorres2016}, Canto Torres/Sendrier develop a framework which includes all known 
variants of information set decoding for generic linear codes, and show that all variants of information set decoding that 
fit within this framework will have asymptotically identical cost when the error rate is sublinear.  
Thus new information set decoding algorithms that fall within this framework will not significantly change the 
security of McEliece and its variants.

\section{The McEliece Cryptosystem}\label{S-McENie}
In this section we recall the general set-up for the McEliece
cryptosystem and discuss its vulnerability for various choices of
underlying codes.  The central feature is a code that comes with an
efficient decoding algorithm.  As before, let $\mathbb{F}=\mathbb{F}_q$. We set
\begin{equation}\label{e-Geff}
  \cG_{q,n,k,t}=\bigg\{(\cC,\cD) \,\bigg|\, \begin{array}{l} \cC \subseteq \mathbb{F}^n, \mbox{dim}(\cC) = k,\, \mbox{dist}(\cC) \geq 2t+1, \\ \cD \mbox{ efficient $t$-error decoding algorithm for }\cC
  \end{array}\bigg\}.
\end{equation}
In this terminology, the public key of the cryptosystem is simply the
code~$\cC$, and the decoding algorithm~$\cD$ is kept secret.
Usually, a code is specified by a generator or parity check matrix and
knowing an efficient decoding algorithm is equivalent to knowing a
particularly structured generator or parity check matrix for the code.
In that case the pair $(\cC,\cD)$ above may be replaced with $G$ or
$H$, where the latter are such a generator or parity check matrix.
The public key will then be a sufficiently general and unstructured
generator or parity check matrix of a related code.  Such a matrix may be obtained by a sufficient
scrambling of the structured matrix.  This is the version described
below.

Let $\GL_n(\mathbb{F}):=\{M\in\mathbb{F}^{n\times n}\mid \det(M)\neq0\}$ be the
general linear group and define
\begin{equation}\label{e-Mon}
  \cM_n:=\{P\in\GL_n(\mathbb{F})\mid P\text{ monomial matrix}\},
\end{equation}
where a square matrix is called \emph{monomial} if every row and
column contains exactly one nonzero element.  In other words, monomial
matrices are exactly the products of permutation matrices and
nonsingular diagonal matrices.  Clearly,~$\cM_n$ is a subgroup of
$\GL_n(\mathbb{F})$.  Monomial matrices can be characterized in the following
way.  For any $P\in\GL_n(\mathbb{F})$ we have
\begin{equation}\label{e-WtPres}
  P\text{ is monomial}\Longleftrightarrow\left\{\begin{array}{ll}\text{the isomorphism: } \mathbb{F}^n\longrightarrow\mathbb{F}^n,\ x\longmapsto xP\\
      \text{preserves the Hamming weight.}
    \end{array}\right\}
\end{equation}
This property plays a central role in code-based cryptosystems because
monomial matrices will act on error vectors and therefore do not
change the weight of such vectors.  This is necessary in the following
algorithm so that the legitimate receiver can indeed decode the
received cyphertext.  We fix a field~$\mathbb{F}$ and parameters $k,n$ and
assume these data are known to the public.

\begin{construction}\textbf{McEliece Cryptosystem~\cite{McE78}}\label{McE}\
  \begin{arabiclist}
  \item Pick a parameter~$t$ and a pair $(\cC, \cD) \in\cG_{q,n,k,t}$.
    Suppose~$\cC$ is given by the generator matrix $G\in\mathbb{F}^{k\times
      n}$, which gives rise to the decoding algorithm~$\cD$ correcting at least $t$ errors.
  \item Pick random matrices $S\in\GL_k(\mathbb{F})$ and $P\in\cM_n$ and set
    $\bar{G}:=SGP^{-1}$.
  \item \textbf{Public Key:} $(\bar{G},t)$.
    \\
    \textbf{Secret Key:} $(S,G,P)$.
  \item \textbf{Plaintext Space:} $\cZ:=\mathbb{F}^k$.
  \item \textbf{Encryption:} Encrypt the plaintext $m\in\cZ$ into
    $c:=m\bar{G}+e$, where $e\in\mathbb{F}^n$ is a randomly chosen vector of
    weight at most~$t$.
  \item \textbf{Decryption:}
    \begin{liste}
    \item Compute $c':=cP=mSG+eP$.
    \item Decode~$c'$ with the decoding algorithm~$\cD$ for $\cC$ and
      denote the output by~$m'$.
    \item Return $m'S^{-1}$.
    \end{liste}
  \end{arabiclist}
\end{construction}

Some comments are in order.  Step~(2) serves to mask the specific
generator matrix~$G$.  Note that left multiplication by~$S$ does not
change the code generated by~$G$, whereas right multiplication by~$P$
only mildly changes the code and usually leaves the type of code
invariant (such as GRS, alternant, or Reed-Muller).  In~(5) it is most
secure to choose $e$ of weight as large as possible in order to
minimize an attacker's success when using information-set decoding.
As for~(6) we know that $\dd(c',\im G)\leq\wt(eP)\leq t$.  Thus the
codeword $mSG$ is actually the unique closest codeword to~$c'$ and
therefore applying the decoding algorithm to~$c'$ will return
$m'=mS$. Hence $m'S^{-1}$ is indeed the plaintext~$m$.

\begin{rem}\label{R-SecEqu}
  A cryptosystem dual to the McEliece system is the Niederreiter
  system~\cite{Nie86}. It is based on a (scrambled) parity check
  matrix.  We refer to~\cite{Nie86} or \cite[Sec.~II.B]{LDW94} for a
  detailed description.  In~\cite{LDW94} Li et al.\ showed that the
  McEliece system is equivalent to the Niederreiter
  system~\cite{Nie86} if based on the same underlying code.
  This implies in particular that if one system can be broken then so
  can the other.  However, the Niederreiter system has actually some
  crucial advantages.  Firstly, it allows the public key to be given
  in systematic form without loss of security, resulting in a
  reduction of key size and a reduction of work factor for the
  encryption process.  Thus, while for the McEliece system over the
  field $\mathbb{F}_{2^m}$ the public key has size $mkn$ bits, it can be
  reduced to $mk(n-k)$ bits for the Niederreiter system.  Secondly,
  the McEliece system is actually more vulnerable than the
  Niederreiter system to a specific type of attack.  Indeed,
  encrypting the same message twice with the McEliece system will most
  likely result in different error vectors being added, and the
  difference of the two cyphertexts provides useful information to an
  attacker who has the chance to use the same message twice.  Finally,
  the Niederreiter system can be used to construct digital signature
  schemes~\cite{CFS01}.
\end{rem}

Let us now discuss some crucial details concerning the security of the
McEliece cryptosystem and its variants.  As discussed above, an
attacker can either attempt to break the system by using general
decoding algorithms, without using any potential special structure of
the random-seeming public code, or can attempt to uncover and exploit
the structure that leads to an efficient decoding algorithm.  The second approach has led to breaks of most proposed variants.  We
briefly outline the most prominent cases and successful attacks.

\begin{rem}\label{R-Insecure}\
  \begin{alphalist}
  \item Reed-Solomon and Generalized Reed-Solomon (GRS) codes were
    proposed as a way to reduce key size in the McEliece system by
    Niederreiter~\cite{Nie86}.  However,
    Sidelnikov/Shestakov~\cite{SiSh92} showed that one can efficiently
    recover a structured parity-check matrix for the Niederreiter
    system from the public matrix.
  \item Berger/Loidreau~\cite{BeLo05} proposed using subcodes of GRS
    codes, but Wieschebrink~\cite{Wie10} extended the
    Sidelnikov/Shestakov attack to break this system for almost all
    parameters.  This attack is the first one based on Schur products,
    to be discussed later in this paper.  For the same codes, Couvreur
    at el.~\cite{CGGOT14} derived a nested sequence of subcodes --
    called a filtration -- that expose their vulnerability.  Again
    this attack makes use of the Schur product.  In particular, it
    leads to an attack of GRS based cryptosystems that is different
    from the one in~\cite{SiSh92}.
  \item Wieschebrink~\cite{Wie10} suggested GRS codes with a certain
    number of random columns inserted.  However, using that shortening
    a GRS code results in a GRS code and comparing the dimensions of
    certain Schur products, Couvreur et al.~\cite{CGGOT14} could
    identify the random positions and recover the structure of the
    secret code.
  \item Sidelnikov~\cite{Si94} proposed binary Reed-Muller codes, but
    that has been proven insecure by Minder/Shokrollahi~\cite{MiSh07}.
  \item Janwa/Moreno~\cite{JaMo96} suggested algebraic-geometric
    codes, but this has been broken by
    Couvreur/Marquez-Corbella/Pellikaan~\cite{CMP14}.  The latter
    derived a $t$-error-correcting pair with the aid of a filtration
    that in turn is based on Schur products.  The same authors also
    extended their attack to cryptosystems based on subcodes of
    AG-codes~\cite{CMP15}.
  \item Monico et al.~\cite{mo00p} proposed LDPC codes for the
    McEliece system.  The code structure was disguised by a sparse
    invertible matrix acting on the low density parity check
    matrix. It was pointed out in the original paper that fairly large
    sizes would be required so that a low density parity check
    structure cannot be recovered.  To avoid this problem medium
    density parity check (MDPC) codes were proposed by Baldi
    et. al. \cite{ba07p}. Quasi-cyclic (QC) LDPC codes are preferred
    over general LDPC for their smaller key size, but some classes of
    QC-LDPC codes and QC-MDPC codes were successfully attacked by
    Otmani et al.~\cite{OTD08} and more recently by Guo et. al.
    \cite{gu16p}.  More general MDPC codes are still promising, as we
    will discuss in Section~\ref{S-SCMDPC}.
  \item Bernstein et al.~\cite{BLP10} suggested wild Goppa codes (that
    is, Goppa codes with Goppa polynomial $\gamma^{q-1}$ for some
    irreducible polynomial~$\gamma$) due to their better
    error-correcting properties.  A filtration-type attack based on
    Schur products has been derived in~\cite{COT14} by Couvreur et al.\ for the case where
    the Goppa polynomial comes from a quadratic extension.  
    An additional attack on wild Goppa McEliece was developed in \cite{FPP2014} by Faugere et al.
  \item Baldi et al.~\cite{BBCRS16} suggested variants of the McEliece
    cryptosystem, including adding a rank-1 matrix to the monomial
    matrix in Step~(2).  This forces an extra round in the decryption
    which in essence amounts to testing all values of the underlying
    field.  In~\cite{CGGOT14} Couvreur et al.\ presented an attack
    based on Schur products.  They determined a specific subspace of
    the public code with codimension one from which the secret code
    can be identified with the aid of Wieschebrink's
    attack~\cite{Wie10}.  One should note that adding a rank-1 matrix
    to the monomial matrix leads to a public code that is not
    monomially equivalent to the secret code anymore, and thus in
    general is not in the same code class.  However, it is exactly
    this rank-1 property that provides too much structure and thus
    aids an attacker.

    Another idea is to replace the permutation matrix in the monomial
transformation with a matrix whose inverse has an average row weight
of $m$ where $1<m<\!< n$. This idea probably goes back to Baldi
et~al.~\cite{BBC08} who studied this idea in the context of McEliece
versions of LDPC codes.  Baldi et al.~\cite{BBCRS16} further elaborated this idea
by using values of $m$ in the range $1<m<2$ in the
disguising of generalized Reed Solomon codes. For these systems
Couvreur et. al.~\cite{co15a} developed a polynomial time attack which
works best when the average row weight $m$ is less than $1+R$ where
$R$ is the code rate.
     A similar idea has been employed before in the context
    of rank-metric codes by Gabidulin et~al.~\cite{ga91p} but as in the
    rank-metric coding setting vulnerabilities were discovered by these
    subcode constructions. See Overbeck~\cite{ov08} for details.
  \end{alphalist}
\end{rem}
\medskip
A remaining class of codes that seems to be efficiently decodable and
at the same time secure for cryptosystems is subfield-subcodes, such
as alternant codes or, more specifically, Goppa codes.  The latter are
in fact the ones originally suggested by McEliece~\cite{McE78}.  The
reason is that for these codes the generator
matrix~$G$ chosen in Step~(1) of Construction~\ref{McE}
is itself already highly unstructured.
Therefore its scrambled version~$\overline{G}$ 
has -- to this day -- fended off any attack for recovering the
original code.
As a consequence, an attacker can only resort to decoding the random
looking public code.  Choosing sufficiently large parameters~$n,k$,
any desired security level can be reached; see~\cite{BLP08} for latest
details on the Goppa code case.  However, the disadvantage of this
choice is the prohibitively large key size.

Many of the above listed attacks, resulting in an error-correcting
pair, amount to computing a nested sequence of subcodes of the public
code based on certain Schur products.  This strategy relies on the
facts that, firstly, the Schur square of GRS and other AG-codes is of
much lower dimension than those of a random code, and, secondly, that
the attacker knows the algebraic type of the secret code.

We now introduce the Schur product and Schur square and further
discuss their importance.

\begin{definition}\label{D-Schur}
  For $x=(x_1,\ldots,x_n),\,y=(y_1,\ldots,y_n)\in\mathbb{F}^n$ define the
  \emph{star product} as $x\ast y:=(x_1y_1,\ldots,x_ny_n)$.  For two
  codes $\cC_1,\,\cC_2\subseteq\mathbb{F}^n$ define the \emph{Schur product}
  as
  \[
  \cC_1\ast\cC_2:=\spann_{\mathbb{F}}\{x*y\mid x\in\cC_1,\,y\in\cC_2\}.
  \]
  We set $\cC^{(2)}:=\cC\ast\cC$ and call it the \emph{Schur square}
  of~$\cC$.
\end{definition}
Clearly, the star product is commutative and bilinear.  Moreover, if
the code $\cC\subseteq\mathbb{F}^n$ has basis $g_1,\ldots,g_{k}$, then
\begin{equation}\label{e-SchurGenSys}
  \cC^{(2)}=\spann_{\mathbb{F}}\{g_i\ast g_j\mid i=1,\ldots,k,\,j=i,\ldots,k\},
\end{equation}
and thus,
$\dim(\cC^{(2)})\leq\min\big\{{\textstyle{k+1\choose2}},\,n\big\}$.
One even has the following generic result.
\begin{prop}[\mbox{\cite[Prop.~2]{MCP12}}]\label{P-SchurDim}
  Let $n>{k+1\choose 2}$ and~$\cC$ be a randomly chosen
  $[n,k]$-code. Then
  \[
  \mbox{\rm{Prob}}\Big(\!\dim(\cC^{(2)})={\textstyle{k+1\choose2}}\Big)=1.
  \]
\end{prop}
However, for a GRS code~$\cC$ the dimension of~$\cC^{(2)}$ is actually
much smaller.  Indeed, a $k$-dimensional GRS code is given by a
generator matrix of the form $G=(\alpha_j^iz_j)\in\mathbb{F}^{k\times n}$ for
distinct elements $\alpha_1,\ldots,\alpha_n\in\mathbb{F}$ and nonzero elements
$z_1,\ldots,z_n\in\mathbb{F}^*$.  Denoting the resulting code by
$\GRS_{n,k}(\alpha,z)$ one easily verifies that
\begin{equation}\label{e-SchurGRS}
  \GRS_{n,k}(\alpha,z)\ast \GRS_{n,k'}(\alpha,z')=\GRS_{n,k+k'-1}(\alpha,z*z').
\end{equation}
Thus we have
\begin{equation}\label{e-SchurDimGRS}
  \dim\big(\GRS_{n,k}(\alpha,z)^{(2)}\big)=\min\{2k-1,\,n\}.
\end{equation}
This almost trivial fact aids an attacker if the cryptosystems are
based on GRS codes.  In order to see this, one first observes that the
Schur square of the secret code and that of the public code are
closely related due to the following trivial property of the star
product:
\begin{equation}\label{e-SchurP}
  \text{for any $x,\,y\in\mathbb{F}^n$ and $P\in\cM_n$ we have }(xP)*(yP)=(x*y)\tilde{P},
\end{equation}
where $\tilde{P}$ is the monomial matrix $\big((P_{i,j})^2\big)$,
i.e., each entry of~$P$ is squared.  As a consequence, using the
bilinearity of~$\ast$ we obtain
$\im(\bar{G})^{(2)}=\im(G)^{(2)}\widetilde{P^{-1}}$ for the data as
in Construction~\ref{McE} (and where again $\widetilde{P^{-1}}$ is the
matrix~$P^{-1}$ with each nonzero entry squared).  This enables an
attacker to make use of the public code in order to compute the Schur
square of the underlying secret code -- up to the monomial
matrix~$\tilde{P}$: by Proposition~\ref{P-SchurDim} and
Identity~\eqref{e-SchurDimGRS} she can immediately distinguish the
public code from a random code.  As mentioned in
Remark~\ref{R-Insecure}, this is often sufficient to successfully
attack the system.  In particular, it breaks systems based on GRS
codes or AG-codes (the latter for a wide set of parameters).

In the next sections we present two new variations of the McEliece
cryptosystem.  The first one uses the idea of generalizing the monomial matrix to
enhance the masking as described in Remark~\ref{R-Insecure}(h) but where we use row weight $m=2$.
The other proposes the use of a special class of
MDPC codes that have reduced key size and are not quasi-cyclic,
thereby avoiding attacks designed for quasi-cyclic codes.

\section{Variation Based on Weight-2 Masking}\label{S-Wt2}
Recall the sets~\eqref{e-Geff} 
and~\eqref{e-Mon}.  We first propose the following variant of the
McEliece cryptosystem.  It differs from the classical
Construction~\ref{McE} in that we replace the set~$\cM_n$ of monomial
matrices by the set
\begin{equation}\label{e-Wn}
  \cW_n:=\{P\in\GL_n(\mathbb{F})\mid \text{each row of~$P$ has weight }2\}.
\end{equation}
Notice that matrices of the form $I_n+P$, where $P\in\cM_n$ are
in~$\cW_n$ with high probability.  In fact, Sage simulations show that
for $q=n=2^8$ this probability is at least 96\,\%, whereas for
$q=n=2^9$ it even goes up to 98\,\%.  This allows one to easily create
matrices in $\cW_n$.  As we will demonstrate, using matrices
from~$\cW_n$ instead of~$\cM_n$ leads to an improved masking of the
secret code at the cost of restricting the size of errors to half the
original correctable error-size.

\begin{rem}
  We note that the results that follow remain valid if we extend the
  set $\cW_n$ to the set of invertible matrices where all rows have
  weight at most~$2$ and almost all of them have weight equal to~$2$.
\end{rem}

\begin{construction}\textbf{Variant of McEliece Cryptosystem -- Weight-2
    Masking}\label{McE2}\
  \begin{arabiclist}
  \item Pick a parameter~$t$ and a pair $(\cC,\cD) \in \cG_{q,n,k,t}$
    and choose a generator matrix $G$ for~$\cC$, which gives rise to
    the decoding algorithm~$\cD$.
  \item Pick random matrices $S\in\GL_k(\mathbb{F})$ and $P\in\cW_n$ and set
    $\bar{G}:=SGP^{-1}$.
  \item \textbf{Public Key:} $(\bar{G},t)$.
    \\
    \textbf{Secret Key:} $(S,G,P)$.
  \item \textbf{Plaintext Space:} $\cZ:=\mathbb{F}^k$.
  \item \textbf{Encryption:} Encrypt the plaintext $m\in\cZ$ into
    $c:=m\bar{G}+e$, where $e\in\mathbb{F}^n$ is a randomly chosen vector such
    that $\wt(e)\leq t/2$.
  \item \textbf{Decryption:}
    \begin{liste}
    \item Compute $c':=cP=mSG+eP$.
    \item Decode~$c'$ using the decoding algorithm~$\cD$ for $\cC$ and
      denote the output by~$m'$.
    \item Return $m'S^{-1}$.
    \end{liste}
  \end{arabiclist}
\end{construction}
Note that, in contrast with the original McEliece system, the
matrix~$P$ is not monomial, but rather in~$\cW_n$.  The definition of
this set implies $\wt(eP)\leq 2\wt(e)$ for any $e\in\mathbb{F}^n$, and thus
the condition on the error vector~$e$ guarantees that $\wt(eP)\leq t$.
Therefore, as for Construction~\ref{McE}, decoding of $c'=mSG+eP$ yields the
correct $m'=mS$.  In order to make information-set decoding harder one
should choose $\wt(e)=t/2$ or at least close to it.

\begin{rem}\label{R-Nie}
  Just as for the classical system, one can easily formulate a
  Niederreiter analog for this variant.  In this case the secret
  parity check matrix $H\in\mathbb{F}^{(n-k)\times n}$ is transformed to the
  public key $\bar{H}:=S^{-1}HP\T$ with $S\in\GL_{n-k}(\mathbb{F})$ and
  $P\in\cW_n$ randomly chosen.  One may actually choose~$S$ such that
  $\bar{H}$ is in systematic form to reduce the public key size.  The
  plaintext space is $\cZ:=\{m\in \mathbb{F}^n\mid \wt(m)\leq t/2\}$ and the
  encryption of $m\in\cZ$ is given by $c\T:=\bar{H} m\T$.  Then one
  easily checks that decryption works in exactly the same way as for
  the Niederreiter classical system.  We omit further details.
\end{rem}

It appears that multiplication by the matrix $P^{-1}$ removes any
identifiable algebraic structure from the public code.  It changes the
code type, since transforming~$G$ into $GP^{-1}$ does not amount to a
permutation and rescaling of the columns. Note that in general the
matrix $P^{-1}$ is not even sparse for $P\in\cW_n$.

The loss of algebraic structure becomes clear when computing the Schur
square $\bar{\cC}^{(2)}$ of the public code $\bar{\cC}=\im(\bar{G})$.
The simple identity~\eqref{e-SchurP} no longer holds, and the Schur
square of the public code appears quite unrelated to the Schur square
of the secret code.  If we use a GRS code with $2k-1<n$, this can also
be confirmed with the dimension of~$\bar{\cC}^{(2)}$.
Identity~\eqref{e-SchurDimGRS} tells us that whenever the dimension
of~$\bar{\cC}^{(2)}$ is bigger than $2k-1$, the public
code~$\bar{\cC}$ is not a GRS code.  Similarly, a sufficiently large
dimension of the Schur square precludes that the public code is an
AG-code (with a certain range of parameters) thanks to
\cite[Prop.~8]{CMP14}.  In fact, the following data suggest that the
Schur square of the public code behaves as that of a random code.

\begin{example}\label{E-Simul}
  Sage simulations for the parameters $q=n$ as in
  Example~\ref{E-Workfactor} below and various values of~$k$ (with 100
  simulations for each chosen set of parameters) show that masking a
  secret GRS code with a random matrix from~$\cW_n$ results in a Schur
  square of the public code with the following properties.
  \begin{arabiclist}
  \item The Schur square of $\im(\bar{G})$ attains maximum dimension,
    i.e., $\min\big\{{k+1\choose2},n\big\}$.
  \item The Schur square of randomly taken subspaces of $\im(\bar{G})$
    attains maximum dimension.
  \item The Schur square of the orthogonal code $\im(\bar{G})^{\perp}$
    attains maximum dimension.
  \end{arabiclist}
  These maximum dimensions occurred in all simulations and thus we
  conjecture that the above properties appear with probability~$1$.
  In other words, with respect to the Schur square, the public code,
  its subcodes and its dual appear to behave like random codes, see
  Proposition~\ref{P-SchurDim}.  
  This is underscored by a recent result by Weger in~\cite[p.45, Cor.~4]{We2016} who proved that for fixed $[n,k]$ the 
  probability that the Schur square of $\im(\bar{G})$, where~$G$ is a generator matrix of an $[n,k]$-GRS code, 
  tends to~$1$ as the field size~$q$ goes to infinity. 
  An analogous statement is true for the Niederreiter system~\cite[p.~49, Cor.~8]{We2016}. 
  Finally, we wish to add that the masking used
  in Baldi et al.~\cite{BBCRS16} using transformations having average
  row weight $m$ with $1<m<2$ transforms a GRS code into a code whose
  Schur square does not have full rank, see \cite{co15a}.
\end{example}

\begin{example}\label{H-Simul}
Let $\mathcal{H}_q$ be the Hermitian curve with affine model given by
\[\mathcal{H}_q:\quad  x^q+x = y^{q+1}.\]  The projective curve $\mathcal{H}_q$ has a single point at infinity, is smooth of genus $g=\frac{q(q-1)}{2}$ and has $q^3+1$ points over $\mathbb{F}_{q^2}$.  This is the largest number of points possible over $\mathbb{F}_{q^2}$ for a curve of this genus under the Weil bound, making $\mathcal{H}_q$ a $\mathbb{F}_{q^2}$-maximal curve.  In fact, $\mathcal{H}_q$ has the largest genus possible for a $\mathbb{F}_{q^2}$-maximal curve. AG-codes on Hermitian curves have been widely studied \cite{LittleSaintsHeegard, Stichtenoth} because these codes may have length up to $n=q^3$ with alphabet $\mathbb{F}_{q^2}$.

Let $\mathcal{H}_q(\mathbb{F}_{q^2})$ be the set of points on $\mathcal{H}_q$ defined over $\mathbb{F}_{q^2}$, and let $P_{\infty}$ denote the unique point at $\infty$.  Let $\mathcal{P}=(P_1,P_2,\dots, P_n)$, where $P_i\in\mathcal{H}_q(\mathbb{F}_{q^2})\setminus\{P_{\infty}\}$ for $1\leq i\leq q^3$.  Let $E$ be the divisor $aP_\infty$, where $a$ is any natural number with $2g-2<a<q^3$.  The one-point AG-code $C_a=C_L(\mathcal{H}_q,\mathcal{P},E)$ (see \cite[Section 8.3]{Stichtenoth} and \cite{CMP14, CMP15}) is a code of length $n=q^3$, dimension $k=a+1-g$, and minimum distance $d\geq q^3-a$ \cite[Proposition 8.3.3]{Stichtenoth}.

The Schur square of the code $C_a$ is exactly the code $C_{2a}$, which has dimension $k^*=\min\{n, 2a+1-g\}$.  Thus if $2a+1-g<n$, a Hermitian code $C_a$ can be distinguished from a random code of equal dimension using the Schur square.

Sage simulations were carried out on masked Hermitian codes for the following parameter sets:
\begin{itemize}
\item The curve $\mathcal{H}_4$, with $a=20$, $a=25$, and $a=30$, where the code $C_a$ is defined over $\mathbb{F}_{16}$, with length $n=64$ in all cases.  The code $C_a$ has dimension $k=15$, $k=20$, and $k=25$ respectively.
\item The curve $\mathcal{H}_5$, with $a=20$, $a=30$, and $a=40$, where the code $C_a$ is defined over $\mathbb{F}_{25}$, with length $n=125$ in all cases.  The code $C_a$ has dimension $k=11$, $k=21$, and $k=31$ respectively.
\end{itemize}
In each case, the code was masked with a random matrix from $\cW_n$ and the dimension of the Schur square of the masked code was computed in 100 trials.  In all cases, the Schur square of $\im(\bar{G})$ attains maximum dimension, i.e., $\min\big\{{k+1\choose2},n\big\}$.   Thus the algebraic structure appears to be well-hidden for Hermitian codes as well.
\end{example}

Summarizing, it appears that the vulnerability of our variants depends
mainly on the success of direct decoding attacks.  We close this
section with the following work factor estimates.

\begin{example}\label{E-Workfactor}
  We consider the proposed variant of the McEliece cryptosystem and
  the Niederreiter dual (see Remark~\ref{R-Nie}) based on GRS codes
  with parameters $[q,n,k]$ as in the table below.  The public key
  size is given in bytes and we estimate the work factor for Stern
  decoding with the aid of~\eqref{e-WFStern}, and where~$r$ is the
  number of errors to be corrected (the work factor for the
  Lee-Brickell algorithm~\cite{LeBr89} is higher in each case and thus
  omitted).  For an $[n,k]$-GRS code we have $t=(n-k)/2$ and thus we
  may add $r=t/2$ errors in the encryption.  The estimate for
  $\tilde{W}^{\text{bin}}_{p,\ell}$ is the minimum over all possible
  values for $p$ and~$\ell$.
  \[
  \begin{array}{|c|c|c|c|c|}
    \hline
    [q,n,k]\phantom{\Big|}&\text{key size McEliece} &\text{key size Niederreiter} & \text{$\tilde{W}^{\text{bin}}_{p,\ell}$ for $r=\lfloor t/2\rfloor$}\\ \hline\hline
    [2^8,\,2^8,\,2^7]& 32,768  & 16,384& 2^{53}\\ \hline
    [2^9,\,450,\,230]& 116,437  & 56,925 & 2^{81}\\ \hline
    [2^9,\,470,\,240]& 126,900  & 62,100 & 2^{83}\\ \hline
    [2^9,\,2^9,\,200]& 115,200  & 70,200& 2^{82}\\ \hline
    [2^9,\,2^9,\,250]& 144,000  & 73,687 
    & 2^{88}\\ \hline
  \end{array}
  \]
  Let us put these numbers into context.  After an optimized attack on
  the classical McEliece system based on binary Goppa codes, Bernstein
  et al.~\cite{BLP08} recommended a binary Goppa code of size at least
  $[n,k]=[1632,1269]$ (based on a Goppa polynomial in $\mathbb{F}_{2^{11}}[x]$
  of degree~$t=33$).  This leads to a key size of $57,580$ bytes if
  used with the Niederreiter system.
  Only this size guarantees an 80-bit security (i.e., a work factor
  for their attack with at least $2^{80}$ binary operations),
  see~\cite{BLP08}.  Note that RSA-1024, having 80-bit security, has a
  key size of $256$~bytes.  It seems likely that a similar
  optimization of Stern's algorithm plus additional software speedups
  as in~\cite{BLP08} reduces the according work factors in above
  table.  More recently, further improvements of Stern's algorithm
  have been presented by Becker et al.~\cite{BJMM12} and May et
  al.~\cite{MMT11}, but as in~\cite{BLP08} they are tailored to binary
  codes.  It is left to future research whether their results can be
  generalized to our case in order to lower the work factor even
  further.  Notwithstanding further reductions, it appears that our
  proposed system compares quite well to the binary
  $[1632,1269]$-Goppa code.
%
\end{example}

\section{A Variation Based on Spatially Coupled MDPC
  Codes}\label{S-SCMDPC}

In this section, we examine the use of a class of spatially coupled
moderate density parity-check (SC-MDPC) codes for the McEliece
cryptosystem. Spatially coupled codes are known for their
capacity-approaching performance and improved decoding thresholds
(i.e. lowest channel quality where error correction is possible)
\cite{ku11}. As noted in Remark \ref{R-Insecure}, low density codes
and some quasi-cyclic codes are vulnerable to attack. While the
decoding performance of general MDPC codes decreases as density
increases, spatially coupled codes typically have the opposite effect
\cite{ku11}. We note further that spatially coupled codes may be
obtained using several methods (e.g., \cite{pu11}, \cite{am16}).
Our focus will be on the construction of SC-MDPC codes using MDPC
convolutional codes, specifically terminated
convolutional codes arising from quasi-cyclic MDPC codes. For background on terminated convolutional codes arising from quasi-cyclic codes, we refer the reader to \cite{le93c, ta87}.
The resulting SC-MDPC code may be represented using a bipartite Tanner graph \cite{kfl01} and
decoded using standard message-passing belief propagation algorithms \cite{pearl88,kfl01}.
This construction approach ensures that the resulting codes are not quasi-cyclic and yet still yield a smaller key
size than general MDPC codes.

Throughout this section let $\mathbb{F}=\mathbb{F}_2$. We will use $\mathbb{F}[D]^{r\times n}$
and $\mathbb{F}(D)^{r\times n}$ to denote the space of $r\times n$ matrices with entries in $\mathbb{F}[D]$ and $\mathbb{F}(D)$, respectively.
As before, $\GL_r(\mathbb{F})$ denotes the group of $r\times r $ invertible matrices with
entries in $\mathbb{F}$.
For a polynomial $f=\sum_{i=0}^n f_i D^i\in\mathbb{F}[D]$ we define the {\sl weight} of~$f$ as number of nonzero coefficients, thus
$\wt(f)=|\{i\mid f_i\neq0\}|$.
Note that $\wt(f+g)\leq\wt(f)+\wt(g)$ and $\wt(fg)\leq\wt(f)\wt(g)$ for all polynomials $f,g$.
Furthermore, we recall the notion of a circulant matrix.
For a vector $a=(a_1,\ldots,a_m)\in\mathbb{F}^m$ denote by $C_a\in\mathbb{F}^{m\times m}$ the circulant matrix
with $i$-th row given by the cyclic shift $(a_i,\ldots,a_m,a_1,\ldots,a_{i-1})$ for $i=1,\ldots,m$.
Moreover, we associate with $a$ and~$C_a$ the polynomial $f_a=\sum_{i=0}^{m-1}a_{i+1}D^i\in\mathbb{F}[D]$.

  Let $H \in \mathbb{F}^{rm\times mn}$ be the parity check matrix of a
  QC-MDPC code, that is, $H=(C_{a_{ij}})_{i=1,\ldots,r\atop j=1,\ldots,n}$, where
  $C_{a_{ij}}\in\mathbb{F}^{m\times m}$ are circulant matrices.
  We associate with it the polynomial matrix $H(D)=(f_{ij})\in\mathbb{F}[D]^{r\times n}$, where
  $f_{ij}:=f_{a_{ij}}$ is the polynomial associated with the circulant $C_{a_{ij}}$.
  Then $H(D)$ is a polynomial parity-check representation of a SC-MDPC code.
  It will represent the private key in the McEliece cryptosystem.
  Let $p,r$ be parameters such that
  $\wt(f_{ij})\leq p$ and $\deg(f_{ij})\leq d$ for all $i,j$.

  Let now $H'(D)=TH(D)$ for some $T\in\GL_r(\mathbb{F})$.
  Then clearly,~$H'(D)$ and~$H(D)$ have the same row space. (It is assumed here that $H(D)$ and $H'(D)$ have rank $r$. This is guaranteed with high probability when the choice of the entries in $H(D)$ are sufficiently random.)
  From the properties of the weight of polynomials it follows
  that $\wt(f)\leq rp$ and $\deg(f)\leq d$ for all entries~$f$ of~$H'(D)$.

\begin{rem}
More generally,
  one can use a non-singular matrix $T(D)\in
  \mathbb{F}[D]^{r\times r}$ having at most $t$ non-zero coefficients in each
  entry to make it harder for an eavesdropper to guess the structure
  of $H(D)$. However, the number of non-zero coefficients in each
  polynomial entry of $H'(D)$ will be larger as a result, increasing the overall key size.
\end{rem}

  We now compute a specific generator matrix for~$H'(D)$ in order to give an upper bound for the key
  size of our public key.
  Recall that $H(D)$ has rank~$r$.
  Thus, without loss of generality we may assume $H'(D)=[C(D)\mid M(D)]$, where $C(D)\in\mathbb{F}[D]^{r\times r}$
  is non-singular and $M(D)$ denotes the last $n-r$ columns of $H'(D)$.
  Then the matrix $H''(D):=C(D)^{-1}H'(D)$ is of the form
  $H''(D)=[I\mid P(D)]$, where $P(D)=\frac{1}{\det(C(D))}\text{Adj}(C(D))M(D)\in \mathbb{F}(D)^{r\times (n-r)}$ and where
  $\text{Adj}(C(D))$ is the classical adjoint of the matrix $C(D)$.
  Define $G'(D):=[P'(D)^T\mid \det(C(D))\cdot I_k]$, where $k:=n-r$ and $P'(D)=\text{Adj}(C(D))M(D)$.
  Then clearly $G'(D)H'(D)^T=0$ and thus $G'(D)$ is a polynomial generator matrix for the code with parity-check matrix $H'(D)$.
  Furthermore, it is easy to see that every entry~$f$ in~$G'(D)$ satisfies
  $\wt(f)\leq r! (rp)^r$  and $\deg(f)\leq d^r$.

  We will use $G'(D)$ for the public key.
  The number of distinct non-zero polynomial entries of~$G'(D)$ is $r(n-r)+1$.
   Since the degree of each entry is at most $d^r$, each non-zero coefficient requires $\log_2(d^r)$ bits to represent it, and thus
  the key size is $(r(n-r)+1)r! (rp)^r r\log_2(d)$ bits.

 Let $d'$ be the actual maximum degree among all entries in $G'(D)$. A message to be encrypted is represented in the form
  $$
  m(D) =(m_1(D), m_2(D),\ldots,m_k(D))\in\mathbb{F}[D]^k,
  $$
  where $\deg(m_i(D))\leq z-1 -d'$ for $i=1,\ldots,k$ for some chosen $z$.
   The message $m(D)$
  is encoded as
  \[
    c(D) = m(D)G'(D) + e(D),
  \]
  where $e(D) =(e_1(D)\dots,e_n(D))\in\mathbb{F}[D]^n$ is an error polynomial with $\deg(e_i(D))\leq z-1$
  for all~$i$ and $\sum_{i=1}^n \wt(e_i(D)) =t$.
  The resulting SC-MDPC code has dimension approximately $k(z-d'-1)$,
  block length $nz$, and rate approximately $\frac{k(z-d'-1)}{nz}$.

  The intended receiver takes an encrypted message $c(D)$ and uses the
  graphical representation of $H(D)$ (i.e., the private key) and the
  belief propagation algorithm to decode $c(D)$. The value of $t$ in
  the choice of the error vector $e(D)$ is such that $t$ is close to
  the maximum value for the belief propagation decoder on $H(D)$ to
  recover the message $m(D)$ with high probability (e.g, probability
  $1 - 10^{-6}$).

\begin{rem}
  Alternatively, one can use $G''(D) = S G'(D)$ as the public key,
  where $S\in\GL_{n-r}(\mathbb{F})$ is a random matrix and is known as part of
  the private key. This will increase the security of the system at
  the cost of an increase in key size.
\end{rem}

We now summarize this variation of the McEliece cryptosystem.

Denote by $ \cH_{n,m,r,z,d,t}$ the set of all pairs $(\cC,\cT)$, where
$\cC\subseteq\mathbb{F}^{mn}$ is a QC-MDPC code with circulant block size~$m$
and associated convolutional SC-MDPC code of dimension $k:=n-r$ over $\mathbb{F}(D)$ and where $\cT$ represents a
Tanner graph of that code that can correct up to~$t$ errors using a message passing decoder with memory
(edge-spreading depth)~$d$.  Here, $z$ determines the length of the terminated SC-MDPC code.

\begin{construction}\textbf{Variant of McEliece Cryptosystem --
    SC-MDPC}\label{McE3}\
  \begin{arabiclist}
  \item Pick a parameter~$t$ and a pair $(\cC, \cT) \in
    \cH_{n,m,r,z,d,t}$. Suppose $\cC$ corresponds to parity-check
    matrix $H(D)\in \mathbb{F}[D]^{r\times n}$, which gives rise to the Tanner
    graph $\cT$.
  \item Pick a random $T\in\GL_r(\mathbb{F})$, and form $H'(D)=TH(D)$.
  Let $G'(D)$ be a corresponding generator matrix as described earlier with maximum degree $d'$ among its entries.
  \item \textbf{Public Key:} $(G'(D),t,z,d')$.
    \\
    \textbf{Secret Key:} $(T, H(D))$.
  \item \textbf{Plaintext Space:}
      $\{(m_1(D), m_2(D), \dots,m_k(D)) \in \mathbb{F}[D]^k\mid \deg(m_i(D))\leq z-1-d'\}$.

  \item \textbf{Encryption:} Encrypt the plaintext $m(D)$ into
    $c(D):=m(D)G'(D)+e(D)$, where $e(D)\in\mathbb{F}_2[D]^n$ is randomly
    chosen such that $\deg(e_i(D))\leq z-1$ for all~$i=1,\ldots,n$ and $\sum_{i=1}^n\wt(e_i(D))\leq t$.
  \item \textbf{Decryption:} Decode $c(D)$ using a belief propagation
    algorithm on the graphical representation of $H(D)$.
    Return the resulting output $m(D)$.
  \end{arabiclist}
\end{construction}

We now assess the security of the SC-MDPC cryptosystem.

\textbf{\emph{Classical decoding attacks:}} Because the published
generator matrix $G'(D)$ is denser than $H(D)$ and hides the underlying structure
of the spatially coupled structure of the code~$\cC$, this system is
not specifically compromised with respect to classical decoding
attacks like the information-set decoding described in \cite{St89}.
 Furthermore, the proposed system is better
than systems utilizing quasi-cyclic codes, since the cyclic shift
property of codewords may be exploited to diminish the complexity of
attacks on the QC-MDPC based cryptosystem \cite{ba16p}. As the SC-MDPC
codes proposed here are no longer QC due to termination, classical
decoding attacks on the SC-MDPC based cryptosystem have a work factor
in the order of $W(nz,k(z-d'-1),t)$, where $W(N,K,w)$
represents the number of elementary operations needed to successfully
search for codewords of weight~$w$ in a code with length~$N$ and
dimension $K$ \cite{BJMM12, ba16p}. Thus, SC-MDPC based cryptosystems are expected to be
better than QC-MDPC based cryptosystems against classical decoding
attacks.

\textbf{\emph{Key recovery attacks (KRA):}} These involve recovering
the private key from the public key and using the private key to
decrypt the ciphertext. Even when the private key is not exactly
recovered, the attack may be successful if an alternate private key is
found that can successfully decode the ciphertext.  In the proposed
SC-MDPC cryptosystem, an attacker can reconstuct $H''(D)$ from the
public key $G'(D)$. $H''(D)$ has row weight upper bounded by
$(n-r+1)r!(rp)^r$. As $H''(D)$ is expected to have inferior error
correction performance compared to the matrix $H(D)$ with message
passing decoding, one approach is to transform $H''(D)$ into a sparse
matrix using matrix transformations. An alternative approach for the
KRA involves guessing the rows of the dual code using the
public key $G'(D)$.  From \cite{ba16p,mi13}, it can be shown
that if the row weight of $H(D)$ is $pn$, then the work factor for KRA
attack on the SC-MDPC system is in the order of
$W(nz,rz-(n-r)(d'+1),np)$.

\begin{example}\label{E-SCMDPC}
  Let $H(D)$ be the $2\times 5$ polynomial matrix
  \[
     {\scriptsize H(D)=\begin{pmatrix}
        1+D+D^{25}+D^{31}+D^{50}\!&\!D^3+D^{30}\!&\!D^3+D^{17}+D^{29}\!&\!D+D^{32}+D^{48}\!&\!1+D^4+D^{39}+D^{50}\\
        D+D^2+D^{30}\!&\!D^3+D^{27}+D^{38}\!&\!D^2\!&\!D^2+D^{23}+D^{37}\!&\!1+D+D^{44}
      \end{pmatrix}}.
  \]
  Then $\wt(f)\leq p:=5$ and $\deg(f)\leq d:=50$ for all entries~$f$ of~$H(D)$.
  Choosing $T=\Smallfourmat{1}{1}{0}{1}$
 and setting $H'(D):=TH(D)$, we obtain the associated generator matrix
  \[G'(D) = \begin{pmatrix}
      a_1(D) & a_2(D) & \Delta(D) & 0 &0\\
      a_3(D) & a_4(D) & 0 &\Delta(D)& 0\\
      a_5(D) & a_6(D) & 0 & 0&\Delta(D)
    \end{pmatrix},
  \]
  where

   \noindent{\scriptsize
    $\Delta(D) = D^3+D^5+D^{27}+D^{31}+D^{32}+D^{33}+D^{34}+D^{38}
    +D^{39}+D^{52}+D^{53}+D^{58}+D^{60}+D^{63}+D^{69}+D^{77}+D^{88}$,\\
    $a_1(D) =D^5+D^6+D^{20}+D^{30}+D^{41}+D^{44}+D^{55}+D^{56}+D^{67} $,\\
    $a_2(D) =D^2+D^3+D^4+D^5+D^{18}+D^{19}+D^{27}+D^{30}+D^{31}+D^{47}+D^{52}+D^{59}$,\\
    $a_3(D) = D^4+D^5+D^{26}+D^{28}+D^{32}+D^{35}+D^{39}
    +D^{40}+D^{51}+D^{53}+D^{59}+D^{67}+D^{70}+D^{75}+D^{86}  $,\\
    $a_4(D) =D^{23}+D^{24}+D^{27}+D^{31}+D^{34}+D^{37}+D^{38}
    +D^{48}+D^{49}+D^{50}+D^{52}+D^{54}+D^{68}+D^{73}+D^{78}+D^{87}   $,\\
    $a_5(D) = D^4+D^7+D^{27}+D^{30}+D^{38}+D^{47}+D^{53}+D^{66}+D^{74}+D^{88}  $,\\
    $a_6(D) =  1+D+D^5+D^6+D^{25}+D^{26}+D^{30}+D^{31}
    +D^{32}+D^{34}+D^{40}+D^{41}+D^{44}+D^{45}+D^{50}+D^{52}+D^{75}+D^{80}+D^{94} $.
  }

  \noindent Thus the entries of~$G'(D)$ have maximal degree
  $d'=94$ (as opposed to the upper bound $d^r=50^2$) and the sum of the weights of all entries is $98$.
  Hence the key size is around $98\cdot \log_2(94)=643$ (as opposed to the upper bound
  $(r(n-r)+1)r!(rp)^r \log_2(d^r)$).
  The final SC-MDPC code with generator matrix $G'(D)$ obtained from
  the above construction uses $z=2000$.
  The overall block length is $N=5(2000) =10000$ bits and dimension $K\sim
  (n-r)z-(n-r)(d'+1)=5715$.
  Using simulations, the value $t=140$
  was found to be a good choice for the SC-MDPC cryptosystem with
  these parameters. Decoding via $H(D)$ using message passing decoder
  is able to correct errors of weight $t$. 
Using the analysis in \cite{ba16p}, we can show that for classical decoding attacks, the cryptosystem in Example 5.5 achieves  $> 2^{80}$ level of security for $t \sim$ 140. However, for  key recovery attacks, the minimum row weight of the matrix $H(D)$ in Example~\ref{E-SCMDPC} is too small to guarantee the same security.
Denser parity check matrices (see Table 2 in \cite{MTSB13})  are necessary to achieve a high level of security against KRA and will result in a larger key size.

  \begin{figure} 
    \centerline{\resizebox{5.5in}{3.6in}{\includegraphics{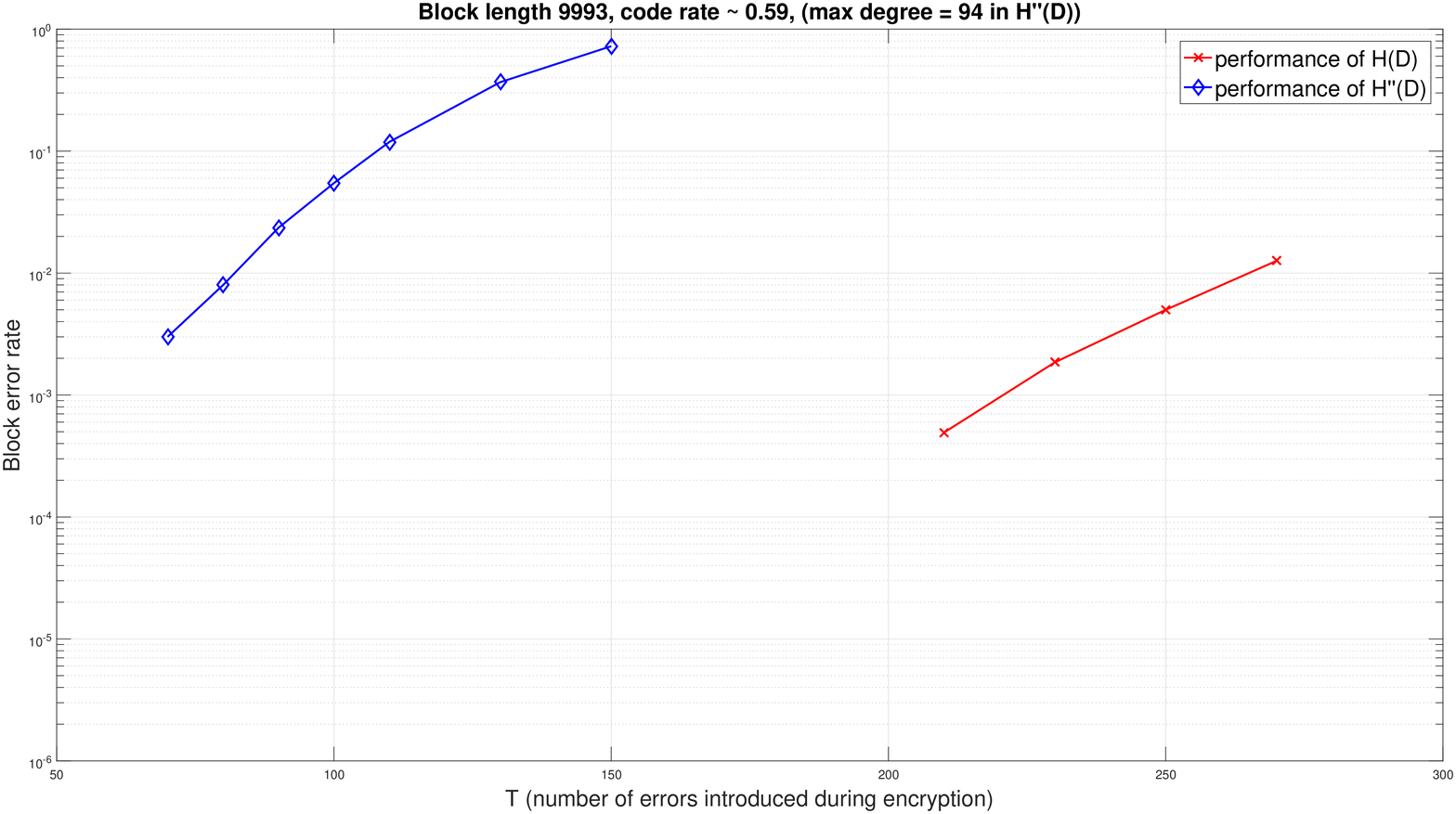}}}
    \caption{Performance of the SC MDPC cryptosystem for both the
      intended receiver using $H(D)$, and an attacker using $H''(D)$.}
    \label{scmpdc_sim}
  \end{figure}

  Figure~\ref{scmpdc_sim} shows the performance of the SC-MDPC
  cryptosystem designed from Example~\ref{E-SCMDPC}. The choice of entries in
  $H(D)$ resulted in some isolated vertices in the Tanner graph of the
  SC-MDPC code that were discarded. Thus, the overall blocklength is
  slightly less than the intended block length of 10,000 bits, and the
  code rate is approximately 0.59. The figure shows that the intended
  receiver can decode with low probability of error using a belief
  propagation decoder when the number of injected errors during
  encryption is $t\le 140$. An attacker who can construct $H''(D)$ has
  probability $\sim$ 1 of failing to decode when the number of bit errors
  injected during encryption is around 140 errors. Although no error floor was observed in the simulation, error floor performance at very low block failure rates of these codes with message passing decoding needs to be investigated further. This will not change the attacker's ability but could make the system less robust for the receiver.

\end{example}

\bf{Acknowledgements}
We would like to thank the organizers of the IPAM workshop on Algebraic Geometry for Coding Theory and Cryptography for inviting us to the event. 
Thanks also go to Mike O'Sullivan for helpful conversations and to the anonymous referee for kind suggestions.
HGL was partially supported by the National Science Foundation Grant DMS-1210061 and by the grant \#422479 from the Simons Foundation.
JB was supported by the US Department of Education GAANN Grant P200A120068.
JR was  partially supported by the Swiss National Science Foundation under grant no. 169510.
BM was partially supported by the National Security Agency under grant H98230-16-1-0300.

\begin{thebibliography}{10}

\bibitem{nist16}
{Report on Post-Quantum Cryptography}.
\newblock Technical report, National Institute of Standards and Technology,
  February 2016.
\newblock NISTIR 8105.

\bibitem{am16}
B.~Amiri, A.~Reisizadehmobarakeh, H.~Esfahanizadeh, J.~Kliewer, and L.~Dolecek.
\newblock Optimized design of finite-length separable circulant-based
  spatially-coupled codes: An absorbing set-based analysis.
\newblock {\em IEEE Trans. on Communications}, 6778(c):1--1, 2016.

\bibitem{BBCRS16}
M.~Baldi, M.~Bianchi, F.~Chiaraluce, J.~Rosenthal, and D.~Schipane.
\newblock Enhanced public key security for the {M}c{E}liece cryptosystem.
\newblock {\em J. Cryptol}, 29:1--27, 2016.
\newblock arXiv: 1108.2462, 2011.

\bibitem{BBC08}
M.~Baldi, M.~Bodrato, and F.~Chiaraluce.
\newblock A new analysis of the {M}c{E}liece cryptosystem based on {QC-LDPC}
  codes.
\newblock In {\em Proc. Security and Cryptography for Networks 2008 (Amalfi,
  Italy), LNCS vol. 5229}, page 246–262, 2008.

\bibitem{ba07p}
M.~Baldi and F.~Chiaraluce.
\newblock Cryptanalysis of a new instance of mceliece cryptosystem based on
  qc-ldpc codes.
\newblock In {\em 2007 IEEE International Symposium on Information Theory},
  pages 2591--2595. IEEE, 2007.

\bibitem{ba16p}
M.~Baldi, P.~Santini, and F.~Chiaraluce.
\newblock Soft {McEliece}: {MDPC} code-based {McE}liece cryptosystems with very
  compact keys through real-valued intentional errors.
\newblock {\em arXiv preprint arXiv:1606.01040}, 2016.

\bibitem{BJMM12}
A.~Becker, A.~Joux, A.~May, and A.~Meurer.
\newblock Decoding random binary linear codes in \mbox{$2^{n/20}$}: {H}ow
  \mbox{$1+1=0$} improves information set decoding.
\newblock In {\em Eurocrypt 2012, LNCS vol. 7237}, pages 520--536, 2012.

\bibitem{BeLo05}
T.~Berger and P.~Loidreau.
\newblock How to mask the structure of codes for cryptographic use.
\newblock {\em Des. Codes Cryptogr.}, 35:63--79, 2005.

\bibitem{BMT78}
E.~Berlekamp, R.~McEliece, and H.~{van Tilborg}.
\newblock On the inherent intractability of certain coding problems.
\newblock {\em IEEE Trans. Inform. Theory}, IT-24:384--386, 1978.

\bibitem{BLP08}
D.~Bernstein, T.~Lange, and C.~Peters.
\newblock Attacking and defending the {M}c{E}liece cryptosystem.
\newblock {\em Post-Quantum Cryptography}, pages 31--46, 2008.

\bibitem{BLP10}
D.~Bernstein, T.~Lange, and C.~Peters.
\newblock Wild {M}c{E}liece.
\newblock In {\em Selected Areas in Cryptography}, pages 143--158, 2010.

\bibitem{CaCh98}
A.~Canteaut and F.~Chabaud.
\newblock A new algorithm for finding minimum-weight words in a linear code:
  application to {M}c{E}liece's cryptosystem and to narrow-sense {BCH} codes of
  length~511.
\newblock {\em IEEE Trans. Inform. Theory}, IT-44:367--378, 1998.

\bibitem{CantoTorres2016}
R.~Canto~Torres and N.~Sendrier.
\newblock Analysis of information set decoding for a sub-linear error weight.
\newblock In Takagi, editor, {\em Post-Quantum Cryptography, LNCS vol.~9606},
  pages 144--161. Springer Cham, 2016.

\bibitem{cdfkms14}
D.~J. {Costello, Jr.}, L.~Dolecek, T.~E. Fuja, J.~, Kliewer, D.~J.~M. Mitchell,
  and R.~Smarandache.
\newblock Spatially coupled sparse codes on graphs: {T}heory and practice.
\newblock {\em IEEE Communications Magazine}, 52(7):168--176, 2014.

\bibitem{CFS01}
N.~Courtois, M.~Finiasz, and N.~Sendrier.
\newblock How to achieve a {M}c{E}liece-based digital signature scheme.
\newblock In {\em “ASIACRYPT 2001, LNCS vol. 2248}, pages 157--174. Springer,
  2001.

\bibitem{CGGOT14}
A.~Couvreur, P.~Gaborit, V.~Gauthier-Uma{\~n}a, A.~Otmani, and J.-P. Tillich.
\newblock Distinguisher-based attacks on public-key cryptosystems using
  {R}eed-{S}olomon codes.
\newblock {\em Des. Codes Cryptogr.}, 73:641--666, 2014.

\bibitem{CMP14}
A.~Couvreur, I.~M\'{a}rquez-Corbella, and R.~Pellikaan.
\newblock A polynomial time attack against algebraic geometry code based public
  key cryptosystems.
\newblock In {\em 2014 IEEE International Symposium on Information Theory},
  pages 1446--1450, 2014.

\bibitem{CMP15}
A.~Couvreur, I.~M\'{a}rquez-Corbella, and R.~Pellikaan.
\newblock Cryptanalysis of public-key cryptosystems that use subcodes of
  algebraic geometry codes.
\newblock In {\em Coding Theory and Applications, {\rm CIM Series in
  Mathematical Sciences 3 (R.~Pinto, P.~Rocha Malonek, P.~Vettori, eds.)}},
  pages 133--140. Springer, 2015.

\bibitem{COT14}
A.~Couvreur, A.~Otmani, and J.-P. Tillich.
\newblock Polynomial time attack on wild {M}c{E}liece over quadratic
  extensions.
\newblock In {\em “Advances in Cryptology — Eurocrypt 2014, LNCS vol.
  8441}, pages 17--39. Springer, 2014.

\bibitem{co15a}
A.~Couvreur, A.~Otmani, J.-P. Tillich, and V.~Gauthier-Uma{\~n}a.
\newblock A polynomial-time attack on the {BBCRS} scheme.
\newblock In {\em Public-key cryptography---{PKC} 2015}, volume 9020 of {\em
  Lecture Notes in Comput. Sci.}, pages 175--193. Springer, Heidelberg, 2015.

\bibitem{FPP2014}
J.-C. Faug{\`e}re, L.~Perret, and F.~de~Portzamparc.
\newblock Algebraic attack against variants of {M}c{E}liece with {G}oppa
  polynomial of a special form.
\newblock In Sarkar and Iwata, editors, {\em Advances in Cryptology. ASIACRYPT
  2014. LNCS, vol.~8873}, pages 21--41. Springer Berlin Heidelberg, 2014.

\bibitem{FiSe09}
M.~Finiasz and N.~Sendrier.
\newblock Security bounds for the design of code-based cryptosystems.
\newblock In {\em “Advances in cryptology — ASIACRYPT 2009, LNCS vol.
  5912}, pages 88--105. Springer, 2009.

\bibitem{ga91p}
E.~Gabidulin, A.~Paramonov, and O.~Tretjakov.
\newblock Ideals over a non-commutative ring and their application in
  cryptology.
\newblock In D.~Davies, editor, {\em Advances in Cryptology, EUROCRYPT'91},
  volume 547 of {\em Lecture Notes in Computer Science}, pages 482--489.
  Springer Berlin Heidelberg, 1991.

\bibitem{gu16p}
Q.~Guo, T.~Johansson, and P.~Stankovski.
\newblock A key recovery attack on {MDPC} with {CCA} security using decoding
  errors.
\newblock Cryptology ePrint Archive, Report 2016/858, 2016.

\bibitem{JaMo96}
H.~Janwa and O.~Moreno.
\newblock Mc{E}liece public key cryptosystems using algebraic-geo\-me\-tric
  codes.
\newblock {\em Des. Codes Cryptogr.}, 8:293--307, 1996.

\bibitem{kfl01}
F.~Kschischang, B.~Frey, and H.-A. Loeligar.
\newblock Factor graphs and the sum-product algorithm.
\newblock {\em IEEE Trans. Inform. Theory}, 47(2):498--519, 2001.

\bibitem{ku11}
S.~Kudekar, T.~J. Richardson, and R.~L. Urbanke.
\newblock Threshold saturation via spatial coupling: why convolutional {LDPC}
  ensembles perform so well over the {BEC}.
\newblock {\em IEEE Trans. Inform. Theory}, 57(2):803--834, 2011.

\bibitem{LeBr89}
P.~Lee and E.~Brickell.
\newblock An observation on the security of {M}c{E}liece's public key
  cryptosystem.
\newblock In {\em “Advances in Cryptology — Eurocrypt 1988, LNCS vol. 330},
  pages 275--280. Springer, 1989.

\bibitem{le93c}
Y.~Levy and D.~J. {Costello Jr.}
\newblock An algebraic approach to constructing convolutional codes from
  quasi-cyclic codes.
\newblock {\em DIMACS Series in Discrete Mathematics and Theoretical Computer
  Science}, 14:189--198, 1993.

\bibitem{LDW94}
Y.~X. Li, R.~H. Deng, and X.~M. Wang.
\newblock On the equivalence of {M}c{E}liece's and niederreiter's public-key
  cryptosystems.
\newblock {\em IEEE Trans. Inform. Theory}, IT-40:271--273, 1994.

\bibitem{LittleSaintsHeegard}
J.~Little, K.~Saints, and C.~Heegard.
\newblock On the structure of {H}ermitian codes.
\newblock {\em Journal of Pure and Applied Algebra}, 121:293--314, 1997.

\bibitem{MCP12}
I.~M\`{a}rquez-Corbella and R.~Pellikaan.
\newblock Error-correcting pairs for a public-key cryptosystem.
\newblock arXiv: 1205.3647, 2012.

\bibitem{MMT11}
A.~May, A.~Meurer, and E.~Thomae.
\newblock Decoding random linear codes in \mbox{$\tilde{\mathcal
  O}(2^{0.054n})$}.
\newblock In {\em “ASIACRYPT 2011, LNCS vol. 7073}, pages 107--124, 2011.

\bibitem{McE78}
R.~McEliece.
\newblock A public-key cryptosystem based on algebraic coding theory.
\newblock In {\em DSN Progress Report}, volume~42, pages 114--116, 1978.

\bibitem{MiSh07}
L.~Minder and A.~Shokrollahi.
\newblock Cryptanalysis of the {S}idelnikov cryptosystem.
\newblock In {\em “Advances in Cryptology — Eurocrypt 2007, LNCS vol.
  4515}, pages 347--360. Springer, 2007.

\bibitem{MTSB13}
R.~Misoczki, J.-P. Tillich, N.~Sendrier, and P.~Baretto.
\newblock {MDPC-McE}liece: {N}ew {M}c{E}liece variants from moderate density
  parity-check codes.
\newblock In {\em 2013 IEEE International Symposium on Information Theory},
  pages 2069--2073, 2013.

\bibitem{mi13}
R.~Misoczki, J.-P. Tillich, N.~Sendrier, and P.~S. Barreto.
\newblock {MDPC-McEliece}: {N}ew {McEliece} variants from moderate density
  parity-check codes.
\newblock In {\em Information Theory Proceedings (ISIT), 2013 IEEE
  International Symposium on}, pages 2069--2073. IEEE, 2013.

\bibitem{mld15}
D.~J.~M. Mitchell, M.~Lentmaier, and D.~J. {Costello, Jr.}
\newblock Spatially coupled {LDPC} codes constructed from protographs.
\newblock {\em IEEE Trans. Inform. Theory}, 61(9):4866--4889, 2015.

\bibitem{mo00p}
C.~Monico, J.~Rosenthal, and A.~Shokrollahi.
\newblock Using low density parity check codes in the {McE}liece cryptosystem.
\newblock In {\em Proceedings of the 2000 IEEE International Symposium on
  Information Theory}, page 215, Sorrento, Italy, 2000.

\bibitem{Nie86}
H.~Niederreiter.
\newblock Knapsack-type cryptosystems and algebraic coding theory.
\newblock {\em Problems of Control and Information Theory}, 15:159--166, 1986.

\bibitem{OTD08}
A.~Otmani, J.~Tillich, and L.~Dallot.
\newblock Cryptanalysis of two {M}c{E}liece cryptosystems based on quasi-cyclic
  codes.
\newblock In {\em Proc. First International Conference on Symbolic Computation
  and Cryptography (SCC 2008), Beijing, China}, 2008.

\bibitem{ov08}
R.~Overbeck.
\newblock Structural attacks for public key cryptosystems based on {G}abidulin
  codes.
\newblock {\em J. Cryptology}, 21(2):280--301, 2008.

\bibitem{pearl88}
J.~Pearl.
\newblock Probabilistic reasoning in intelligent systems: Networks of plausible
  inference.
\newblock {\em San Francisco, CA: Morgan Kaufmann, 2nd edition}, ISBN
  1-55860-479-0., 1998.

\bibitem{Pe10}
C.~Peters.
\newblock Information-set decoding for linear codes over \mbox{$\mathbb{F}_q$}.
\newblock In {\em PQCrypto 2010, LNCS vol.\ 6061}, pages 81--94. Springer,
  2010.

\bibitem{pu11}
A.~E. Pusane, R.~Smarandache, P.~O. Vontobel, and D.~J. Costello, {Jr.}
\newblock Deriving good {LDPC} convolutional codes from {LDPC} block codes.
\newblock {\em IEEE Trans. Inform. Theory}, 57(2):835--857, 2011.

\bibitem{Si94}
V.~M. Sidelnikov.
\newblock A public-key cryptosystem based on binary {R}eed-{M}uller codes.
\newblock {\em Discrete Math Appl.}, 4:191--208, 1994.

\bibitem{SiSh92}
V.~M. Sidelnikov and S.~O. Shestakov.
\newblock On insecurity of cryptosystems based on generalized {R}eed-{S}olomon
  codes.
\newblock {\em Discrete Math Appl.}, 2:439--444, 1992.

\bibitem{St89}
J.~Stern.
\newblock A method for finding codewords of small weight.
\newblock In {\em Coding Theory and Applications, LNCS vol.\ 388}, pages
  106--113. Springer, 1989.

\bibitem{Stichtenoth}
H.~Stichtenoth.
\newblock {\em Algebraic Function Fields and Codes}.
\newblock Springer Publishing Company, Incorporated, 2nd edition, 2008.

\bibitem{ta87}
R.~M. Tanner.
\newblock Convolutional codes from quasi-cyclic codes: A link between the
  theories of block and convolutional codes.
\newblock University of California, Santa Cruz, Tech Report UCSC-CRL-87-21,
  Nov. 1987.

\bibitem{We2016}
V.~Weger.
\newblock A code-based cryptosystem using {GRS} codes.
\newblock Master Thesis at the University of Z{\"u}rich (Switzerland), 2016.

\bibitem{Wie10}
C.~Wieschebrink.
\newblock Cryptanalysis of the {N}iederreiter public key scheme based on {GRS}
  subcodes.
\newblock {\em Post-Quantum Cryptography}, pages 61--72, 2010.

\end{thebibliography}
\end{document}